\documentclass[11pt]{cernrep}
\usepackage{epsfig}
\usepackage{cite,./mcite}

\newcommand{\as}{\alpha_s}

\newcommand{\tw}{\textwidth}

\newcommand{\cC}{{\cal C}}
\newcommand{\cE}{{\cal E}}
\newcommand{\cR}{{\cal R}}
\newcommand{\barcC}{{\bar {\cal C}}}

\begin{document}
\title{Resummation}
\author{A.~Banfi$^{1}$, G.~Corcella$^{2}$, M.~Dasgupta$^{3}$, Y.~Delenda$^{3}$, 
G.P.~Salam$^{4}$ and G.~Zanderighi$^{5}$.}
\institute{$^1$ University of Cambridge, Madingley Road, Cambridge CB3 0HE ,U.K\\
           $^2$ Department of Physics, Theory Division, CH-1211 Geneva 23, Switzerland.\\
$^3$ University of Manchester, Oxford Road, Manchester, M13 9PL, U.K \\
$^4$ LPTHE, Universities of Paris VI and VII and CNRS, 75005, Paris, France.\\
$^5$ Fermilab, Batavia, IL 60510, USA.}
\maketitle
\begin{abstract}
We review the work discussed and developed under the topic ``Resummation'' at 
Working Group 2 ``Multijet final states and energy flow'' , of the HERA-LHC Workshop. We emphasise the role played by HERA 
observables in the development of resummation tools via, for instance, 
the discovery and resummation of non-global logarithms.
We describe the event-shapes subsequently  
developed for hadron colliders and present resummed predictions for the same using the automated resummation program CAESAR. 
We also point to ongoing studies at HERA which can be of benefit for future measurements at hadron colliders such as the LHC, 
specifically 
dijet $E_t$ and angular spectra and the transverse momentum of the Breit current hemisphere. 
\end{abstract}

\section{Introduction}
\label{sec:intro}
Resummed calculations are an invaluable tool, both for 
the understanding of perturbative QCD dynamics at all orders 
as 
well as for extracting, as accurately as possible, 
QCD parameters such as the strong coupling, quark masses 
and parton distribution functions. 
These parameters, which cannot be directly 
computed from QCD perturbation theory itself, 
will be vital inputs in new physics searches 
at the LHC. 
Moreover, resummed expressions are also an 
important stepping stone to probing observable distributions 
in regions where non-perturbative power corrections make a significant contribution. In this region one may expect a smearing of the resummed perturbative result with a non-perturbative function (for which one can adopt, for example, a renormalon-inspired model), and the resulting spectrum can be confronted with data to test our understanding of non-perturbative dynamics.  In all these aspects, 
HERA data and 
observables 
have played an important role 
(sometimes significantly underrated  in the literature) 
in furthering our knowledge, without which accurate studies of several observables 
at the LHC would simply not be possible.

 A concrete example of HERA's important role in this regard is the case of event shape 
distributions \cite{H1,*ZEUS}, theoretical studies of which led to the finding of non-global single-logarithmic 
\cite{Dassal1,*Dassal2} effects (discussed in more detail below). Prior to these studies it was widely believed that the HERA distributions, measured in the current hemisphere Breit frame, were trivially related to their $e^{+}e^{-}$ counterparts. Had such 
ideas, based on independent soft gluon emission by the hard partons, 
 been applied directly to similar variables at the LHC,  such as energy flows away from jets, 
the accuracy of theoretical predictions would have been severely compromised leading almost certainly to  erroneous claims and conclusions.

Another area where HERA has played a vital role is in the testing of renormalon inspired models for power corrections, most significantly the dispersive approach  \cite{DMW} to $1/Q$ power corrections, 
 tested against HERA event-shape distributions  and mean-values 
 \cite{DasWeb97}. The fact that HERA data seem to confirm such models
 , where one can think 
 of the power corrections as arising from the emission of a gluon with transverse momentum ${\mathcal{O}} ({\Lambda_\mathrm{{QCD}}})$, is significant for the LHC. This is because the agreement of the renormalon model with data demonstrates that the presence of initial state protons does not affect significantly the form of $1/Q$ corrections. It thus sets limits on the additional non-perturbative contribution that may potentially be generated by the flight of struck partons through the proton cloud, which therefore does not appear to be significant. Once again it is accurate resummed predictions \cite{Dassal3} which have allowed us access to the non-perturbative domain hence strengthening our understanding of power corrections.
 
 One important aspect of resummed studies, till date, is that stringent comparisons of next-to--leading logarithmic resummed predictions with data have only been carried out in cases involving observables that vanish in the limit of two hard partons. Prominent examples reflecting the success of this program are provided by $e^{+}e^{-} \to $ 2 jet event shapes  and DIS (1+1) jet event shapes as well as Drell-Yan vector boson transverse momentum spectra at hadron colliders.
 At the LHC (and hadron colliders in general)  one already has two hard incoming partons and any observable dealing with final state jet production would take us beyond the tested two hard parton situation. Thus dijet event shapes at hadron colliders (discussed in detail later), which involve much more complicated considerations as far as the resummation goes, represent a situation where NLL resummations and power corrections are as yet untested. Bearing in mind the hadronic activity due to the underlying event at hadron colliders, it is important to test the picture of resummations and power corrections for these multiparton event shapes in cleaner environments. Thus LEP three-jet event shapes and similar 
$2+1$ jet event shapes at HERA become important to study in conjunction with 
 looking at resummation of event shapes at hadron colliders.

Predictions for several 
LEP and HERA three-jet event shapes already exist (see e.g 
\cite{Kout,*Koutdis} and for a full list of variables studied 
Ref.~ \cite{dassalrev}) and at this workshop a prominent development 
presented 
was the proposal of several dijet event-shapes in hadron-hadron collisions and 
the resummed predictions for their distributions \cite{BSZhh}.

Existing HERA data can also be usefully employed to study soft gluon radiation dynamics from multi-hard--parton ensembles, in the study of dijet $E_t$ and angular spectra. 
These quantities are somewhat different from event shapes since 
one defines observables based on aggregate jet-momenta and angles 
rather than directly constructing them 
from final-state hadron momenta. Examples are the transverse energy, $E_t$, mismatch between the leading $E_t$ jets in dijet production and the azimuthal correlation between jets $\phi_{jj}$, once again refering to the highest $E_t$ jets in dijet production. For the former quantity there are no direct experimental data as yet, but it is simply related to the dijet total rate in the region of symmetric $E_t$ cuts for which data does exist . 
For the latter quantity similarly 
there are direct experimental data \cite{Zeus1,*Zeus2}. 
These observables have smaller hadronisation corrections scaling as $1/Q^2$ rather than $1/Q$ as for 
most event shapes. They thus offer a good opportunity to test the NLL perturbative predictions alone without necessarily probing non-perturbative effects at the same time \footnote{Although effects to do with intrinsic $k_t$ will eventually have to be accounted for similar to the case of Drell-Yan vector boson $p_t$ spectra.}. 

At this workshop developments were reported on extending existing calculations \cite{BanDas} for cone dijets, 
to different jet algorithms, such as the $k_t$ algorithm, 
comparing to fixed order estimates and performing the leading 
order matching. Once the HERA data has been well described 
similar studies can be carried out for hadron--hadron dijets. 
In fact predictions already exist for hadron-hadron dijet masses near threshold \cite{KOS} but 
are not in a form conducive to direct comparisons with data containing neither 
the jet algorithms in the form 
actually employed in experiment, nor the matching to fixed order. However these calculations provided 
a useful starting point for the calculations presented here, which should eventually lead to direct comparisons with data.

Another area where HERA may play an important role is to establish whether unaccounted for small $x$ effects may be significant in comparing theoretical resummations for e.g. vector boson $p_t$ spectra with experimental data. 
It has been suggested that a 
non-perturbative intrinsic $k_t$, growing steeply with $x$, is required to accomodate 
HERA data for semi-inclusive DIS processes \cite{Yuan1}. 
When this observation is 
extrapolated to the LHC kinematical region there is apparently 
significant small $x$ broadening in the vector boson $p_t$ distribution. Similar effects may well arise in the case of the Higgs boson too. 
However DIS event shape studies in the Breit current hemisphere 
\cite{Dassal3} apparently 
do not acquire such corrections since they are well described by conventional 
NLL resummations supported by dispersive power corrections \cite{DasWeb97}, 
which are $x$ independent \footnote{An exception is the jet broadening \cite{Dassalb} but the $x$ dependence there is of an entirely different origin and nature.}. However there are some important caveats:
\begin{itemize}
\item Unlike vector boson $p_t$ spectra, event shapes receive $1/Q$ hadronisation 
corrections 
      unrelated to intrinsic $k_t$. These could mask $1/Q^2$ terms originating from intrinsic $k_t$ which may yet contain the $x$ dependence in question.
\item It has already been observed that including 
H1 
data for $Q < 30$ GeV does spoil somewhat the agreement with the dispersive prediction of universal power corrections to event shapes \cite{Dassal3}. The origin of this effect could well be extra non-perturbative $k_t$ broadening related to the effects described above for vector boson $p_t$.
\end{itemize}

To get to the heart of this matter a useful variable that has been suggested (see plenary talk by G.~Salam at the first meeting of this workshop) 
is 
the modulus of the vector transverse momentum $\sum_{i\in H_c} \vec{k}_{t,i}$ 
of the current hemisphere in the DIS Breit frame. This quantity is simply related to the Drell-Yan $p_t$ spectra and comparing theoretical predictions, presented here, with data from HERA should help to finalise whether additional small-$x$ enhanced non-perturbative terms are needed to accomodate the data. 
We begin by first describing the results for hadron-hadron event shape variables, discussed by G.~Salam at this workshop. Then we describe 
the progress in studying dijet $E_t$ and angular spectra 
(presented by M.~Dasgupta and G.~Corcella at the working group meetings). Finally we mention the results obtained thus far, 
for the $Q_t$ distribution of the current hemisphere and end with a look at prospects for continuing phenomenology at HERA, that would be of direct 
relevance to the LHC.

\section{Event shapes for hadron colliders}
Event shape distributions at hadron colliders, as has been the case 
at LEP and HERA, 
are important 
collinear and infrared safe quantities, that can be used as tools 
for the extraction of QCD parameters, for instance $\alpha_s$, 
by comparing theory and data. 
In contrast however to more inclusive sources of the same information 
(e.g the ratio of 3 jet to 2 jet rates), 
event shape distributions provide a wealth of other information, some of which 
ought to be crucial in disentangling and 
further understanding the different physics effects, relevant at hadron colliders. These range from fixed-order predictions to resummations, 
hadronisation corrections and, in conjunction with more detailed studies 
assesing the structure of, and role played by, 
the underlying event (beam fragmentation). 

Until recently there have only been limited experimental 
studies of jet-shapes at 
hadron colliders \cite{transthru} and no resummed 
theoretical predictions for 
dijet shape variables at hadron colliders. Rapid recent developments 
(see Ref.~\cite{BSZhh} and references therein) in the field of perturbative 
resummations have now made theoretical estimates possible for a number of such distributions, introduced in \cite{BSZhh} 
which we report on below. 

The three 
main theoretical 
developments that have led to the studies of Ref.~\cite{BSZhh} are:
\begin{itemize}
\item Resummation for hadron-hadron dijet observables depends on describing multiple soft gluon emission from a system of four hard partons. The colour structure of the resulting soft anomalous dimensions is highly non-trivial and was explicitly computed by the Stony Brook group in a series of papers (see e.g 
\cite{KOS} and references therin).

\item The discovery of non-global observables \cite{Dassal1,*Dassal2}. 
The realisation that standard resummation techniques based on angular ordering/independent-emission of soft gluons by the hard-parton ensemble, are not 
valid for observables that are sensitive to emissions in 
a limited angular range, has led to the introduction of observables that are 
made global by construction. This means that one can apply the technology developed by the Stony-Brook group to obtain accurate NLL predictions for these observables, without having to resort to large $N_c$ approximations.

\item The advent of automated resummation \cite{BSZc}. The development of 
generalised resummation formulae and 
powerful numerical methods to determine the parameters and compute the 
functions thereof, 
has made it possible to study several variables at once rather than 
having to perform copious, and in some cases previously unfeasible,  
calculations for each separate observable. 
\end{itemize}
 
We now discuss the different types of variables defined and resummed in \cite{BSZhh}. The first issue one has to deal with is the fact that experimental detectors have a limited rapidity range, which can be modeled by a cut around the beam direction.

\begin{figure}
\begin{center}
\label{cut}
\epsfig{file=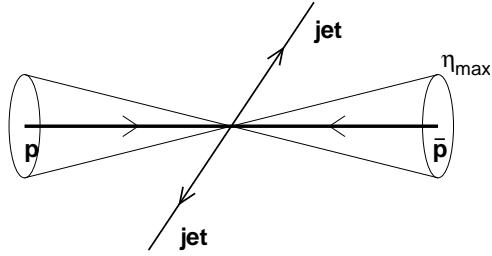, width=0.4\textwidth}
\caption{Cut around the beam direction beyond rapidity $\eta_{\rm{max}}$ corresponding to the maximum rapidity reach of the detectors.}
\end{center}
\end{figure}

This cut would then correspond to a position in rapidity of the edge of 
the most forward detector with momentum or energy resolution and the relevant 
values of the maximum rapidity for measurements is $3.5$ units at the Tevatron and $5$ units at the LHC. One may then worry about gluon emissions beyond 
this rapidity (i.e. inside the beam cut, see Fig.~\ref{cut}) 
that emit softer gluons into the allowed rapidity range, outside the cones depicted in Fig.~\ref{cut}. Such a configuration would of course render the observable non-global.

To get around this potential problem, one can employ an idea 
suggested for 3-jet observables such as out-of--plane momentum flows 
in hadron-hadron collisions \cite{Kouthh}, which helps 
side-step the issue of non-globalness. 
We note that all the observables studied here 
have the following functional dependence on a soft emission, $k$, collinear to 
a given hard leg
\footnote{In general the values of parameters $d,a,b$ and the function $g$ 
depend on the observable considered. 
For more details and constraints on the various parameters that ensure globalness and infrared and collinear safety etc., 
see Ref.~\cite{BSZc}.}
(common to all event shapes studied here and in other processes)
\begin{equation}
\label{form}
V\left({\tilde{p}},k\right) = d \left(\frac{k_t}{Q}\right)^{a} e^{-b \eta}g(\phi),
\end{equation}
where $k_t$ , $\eta$ and $\phi$ are measured wrt a given hard leg and ${\tilde{p}}$ represent the set of hard parton momenta including recoil against $k$ while $Q$ is the hard-scale of the process. 
We are particularly 
interested in emissions soft and collinear to the beam (incoming) partons. 
Then an emission beyond the maximum detector rapidity $\eta \geq \eta_{{\max}}$ corresponds to at most a 
contribution to the observable $V \sim e^{-(a+b_{\mathrm{min}}) \eta_{\mathrm{max}}}$ with $ b_{\mathrm{min}} ={\mathrm{min}}(b_1,b_2)$ and $b_1$ and $b_2$ 
are the values of $b$ associated with collinear emission near beam-partons 1 and 2. 

If one then choses to study the observable over a range of values such that 
\begin{equation}
\label{range}
L \leq (a+b_{\mathrm{min}}) \eta_{\mathrm{max}}, \, \, L \equiv \ln 1/V,
\end{equation}
then emissions more forward than $\eta_{\mathrm{max}}$ do not 
affect the 
observable in the measured range of values. One can thus include the 
negligible contribution from this region and do the calculation {\emph{as if the observable were global}}, ignoring the cut around the beam. Including the 
region beyond $\eta_{\mathrm{max}}$ does not alter the NLL resummed result in the suitably selected range Eq.~\ref{range}.

The price one has to pay is to limit the range of the study of the observable $V$, such that emissions beyond $\eta_{\mathrm{max}}$ make a negligible contribution. As we will mention later this is a more significant restriction for some 
variables compared to others (depending on the parameters $a$ and $b$) but a range of study can always be found over which the observable can be treated as global.
\subsection{Global event shapes}
\begin{figure}
\begin{center}

\epsfig{file=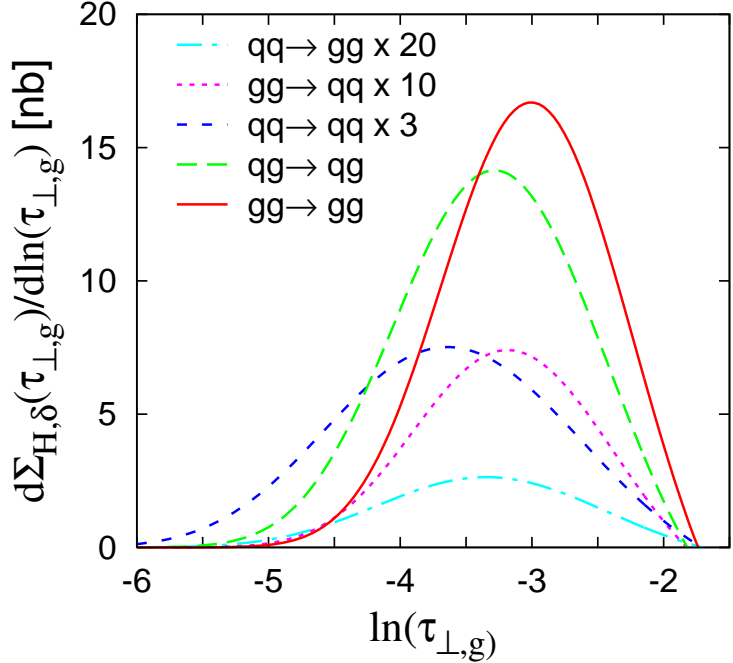,width=0.6 \textwidth}
\caption{The global transverse thrust distribution with the contribution from 
different partonic channels explicitly displayed.}
\label{TG}
\end{center}
\end{figure}

 With the above caveat in place several variables can be safely studied 
(treated as global) over a wide range of values. An explicit example is the 
{\emph{global transverse thrust}} defined as:
\begin{equation}
    \label{eq:Ttg}
    T_{\perp,g} \equiv \max_{\vec n_T} \frac{\sum_i |{\vec
        q}_{\perp i}\cdot {\vec
        n_T}|}{\sum_i q_{\perp i}}\,,\qquad \tau_{\perp,g} = 1 - T_{\perp,g}\,,
 \end{equation}
where the thrust axis $\vec{n}_T$ is defined 
in the plane transverse to the beam axis.  
The probability $P(v)$, that the event shape is smaller than some value $v$ behaves as:
\begin{equation}
\label{Prob}
P(v) = \exp{\left [-G_{12} \frac{\alpha_s}{2\pi} L^2+\cdots \right]}, \,\, L=\ln 1/v,
\end{equation}
with $G_{12} = 2 C_B+C_J$, where $C_B$ and $C_J$ represent the total colour charges of the beam and jet (outgoing) partons. The above represents just the 
double-logarithmic contribution. The full result with control of up to next-to--leading single-logarithms in the exponent is considerably more complicated. It contains both 
the Stony-Brook colour evolution matrices as well as multiple emission effects 
(generated by phase-space factorisation). The automated resummation program
CAESAR \cite{BSZc} is used to generate the NLL resummed result shown in Fig.~\ref{TG}.
In this particular case the effect of the cut around the beam direction can 
be ignored for values $\tau_{\perp,g} \geq 0.15 e^{-\eta_{\mathrm{max}}}$. 
We note that it is advisable to leave a safety margin between this value and the values included in measurement.

Other global variables studied 
include the {\emph{global thrust minor}} and the three jet-resolution 
threshold parameter $y_{23}$. For detailed definitions and 
studies of these variables, the reader is refered to \cite{BSZhh}.

We shall now proceed to look at two different ways of defining event shapes in a given central region, which on its own would lead to non-globalness, 
and then adding terms that render them global.
\subsection{Forward suppressed observables}

  \begin{figure}
   \begin{center}
  \epsfig{file=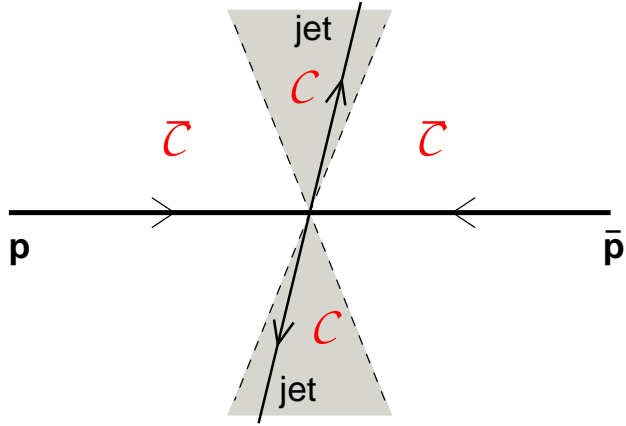}
\label{cones}
\caption{Figure depicting the central region marked $\cC$ , 
containing the two hard jets.}
\end{center}  
\end{figure}

Here we shall examine event shapes defined in a chosen central region $\cC$ 
well 
away from the forward detector edges.
 
First we define central $\perp$ momentum, and rapidity:
    \begin{equation}
      \label{eq:etabar}
      Q_{\perp,\cC} = \sum_{i\in \cC} q_{\perp i}\,,\quad
      \eta_\cC = \frac{1}{
        Q_{\perp,\cC}} \sum_{i\in\cC} \eta_i\, q_{\perp i}\,
    \end{equation}
    and an {\emph{exponentially suppressed forward term,}}
    \begin{equation}
      \cE_\barcC = \frac{1 }{Q_{\perp,\cC}}
      \sum_{i \notin \cC} q_{\perp i} \,e^{-|\eta_i - \eta_\cC|}\,.
    \end{equation} 
 
Then we can define an event shape in the central region $\cC$\footnote{There is considerable freedom on the choice of the central region. For instance this could be a region explicitly delimited in rapidity or the two hard jets themselves.} which on its own would be non-global since we measure emissions just in $\cC$. The addition of 
$\cE_\barcC$ to the event-shape renders the observable 
global as this term includes suitably the effect of emissions in the remaining region $\barcC$. 
The exponential suppression of the added term reduces sensitivity to emissions in the forward region which in turn 
reduces the effect of the beam cut $\eta_{\mathrm{max}}$ considerably, 
pushing its impact to values of the observable where the 
shape cross-section is highly suppressed and thus 
too small to be of interest. 

The event shapes are constructed as described stepwise below: 
 \begin{itemize}
  \item Split $\cC$ into two pieces: {\emph{Up, Down}}
  \item Define {\emph{jet masses}} for each
    \begin{equation}
      \label{eq:mass-XC}
      \rho_{X,\cC} \equiv \frac{1}{Q_{\perp,\cC}^2}
      \Big(\sum_{i\in \cC_X} q_{i}\Big)^2\,,\qquad X = U, D\,.
    \end{equation}
    Define sum and heavy-jet masses
    \begin{equation}
      \label{eq:mass-C-sum-heavy}
      \rho_{S,\cC} \equiv \rho_{U,\cC} + \rho_{D,\cC}\,,\qquad\quad
      \rho_{H,\cC} \equiv \max\{\rho_{U,\cC}, \rho_{D,\cC}\}\,.
    \end{equation}
    Define global extension, with extra forward-suppressed term
    \begin{equation}
      \label{eq:mass-E-sum-heavy}
      \rho_{S,\cE} \equiv \rho_{S,\cC} + \cE_{\barcC}\,,\qquad\quad
      \rho_{H,\cE} \equiv \rho_{H,\cC} + \cE_{\barcC}\,.
    \end{equation}
  \item Similarly: {\emph{total and wide jet-broadenings}}
    \begin{equation}
      \label{eq:BTWE}
      B_{T,\cE} \equiv B_{T,\cC} + \cE_{\barcC}\,,\qquad
      B_{W,\cE} \equiv B_{W,\cC} + \cE_{\barcC}\,.
    \end{equation}
  \end{itemize}

\begin{figure}
\begin{center}
\epsfig{file=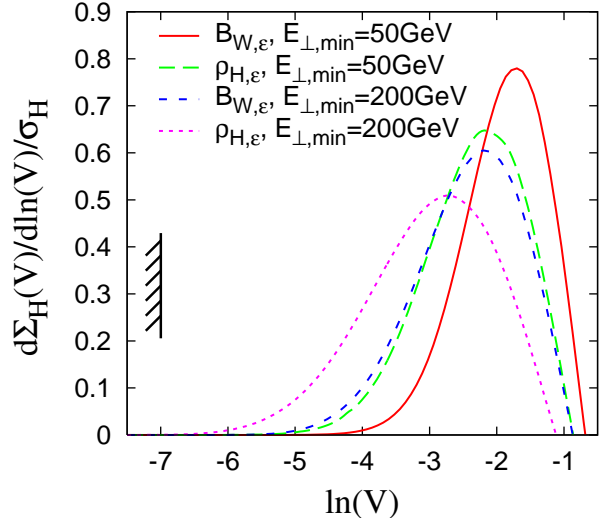}

\caption{NLL resummed predictions from CAESAR for the heavy jet-mass and the wide jet-broadening with the minimum jet transverse energy 
$E_{\perp,{\rm{\min}}}$ values of 50 and 200 GeV as shown.}
\label{fsupp}
  \end{center}
\end{figure}

At the double-log level the results assume an identical form to Eq.~\ref{Prob} with $G_{12}$ representing a combination of total incoming (beam) and outgoing (jet) parton colour charges \cite{BSZhh}. The full NLL resummed results have a substantially more complex form and results from CAESAR \cite{BSZc} are plotted in Fig.~\ref{fsupp}.

\subsection{Indirectly global recoil observables}
Here we study observables that are defined exclusively in terms of particles in the central region but are global. Such observables are already 
familiar from HERA studies. 
As an example, although the current-jet broadening wrt the 
photon axis of the DIS 
Breit frame involves only particles that enter the 
current hemisphere, the current quark acquires transverse momentum by {\emph{recoil}} against 
remnant hemisphere particles. This recoil means that the observable is indirectly sensitive to emissions in the remnant hemisphere which makes the 
observables global. 

To construct similar observables in the hadron-hadron case we observe that by momentum conservation, the following relation holds :
 \begin{equation}
  \sum_{i \in \cC} {\vec q}_{\perp i} = - \sum_{i \notin \cC}
  {\vec q}_{\perp i} 
  \end{equation}
which relates the sum of transverse momenta in $\cC$ to that in the complementary region. Then the central particles can be used to define a recoil term:
\begin{equation}
    \label{eq:recoil-term}
    \cR_{\perp,\cC} \equiv \frac{1}{Q_{\perp,\cC}} \left|\sum_{i
        \in \cC} {\vec 
        q}_{\perp i}\right|\,,
  \end{equation}
which contains an indirect dependence on non-central emissions.

Now we can define event shapes explicitly in terms of central particle momenta 
in $\cC$. Examples are the recoil jet-masses and broadenings

\begin{equation}
    \label{eq:BT}
    \rho_{X,\cR} \equiv \rho_{X,\cC} + \cR_{\perp,\cC}\,,\qquad
    B_{X,\cR} \equiv B_{X,\cC} + \cR_{\perp,\cC}\,, \ldots
  \end{equation}

It is clear that since these observables are defined in terms of central particles alone, the cut around the beam direction is not an issue here. There is however another potential problem. Due to the addition of the recoil term we lose direct exponentiation of the result in variable space. Exponentiation to NLL accuracy only holds in impact-parameter or $b$ space . 

The physical effect in question here is similar to Drell-Yan $Q_T$ spectra where there are two competing mechanisms that lead to a given small $Q_T$, Sudakov suppression of soft emissions and vectorial cancellation between harder emissions. 
Where the latter effect takes over (typically in the region where single-logs are large $\alpha_s L \sim 1$) we get a breakdown of the Sudakov result generated by CAESAR. This result is of the general form:
\begin{equation}
P(V) =e^{Lg_1(\alpha_sL)+g_2(\alpha_sL)+\cdots}.
\end{equation}
The result for recoil observables produced by CAESAR will contain a divergence 
in the single-log function $g_2$ and is cut before the divergence. 
Again for some variables this cut is at a position that significantly reduces 
the range of possible phenomenological studies. 
For other variables the divergence is at values of the observable that are sufficiently small so that only a few percent of the cross-section is beyond the cutoff. An example of the former is the recoil transverse thrust where 15\% of the cross-section lies beyond the cut-off. 
For the recoil thrust minor, in contrast, 
the cutoff has only a moderate effect and much less of the cross-section is cutoff, due to the divergence in $g_2$.
\begin{figure}
\begin{center} 
\epsfig{file=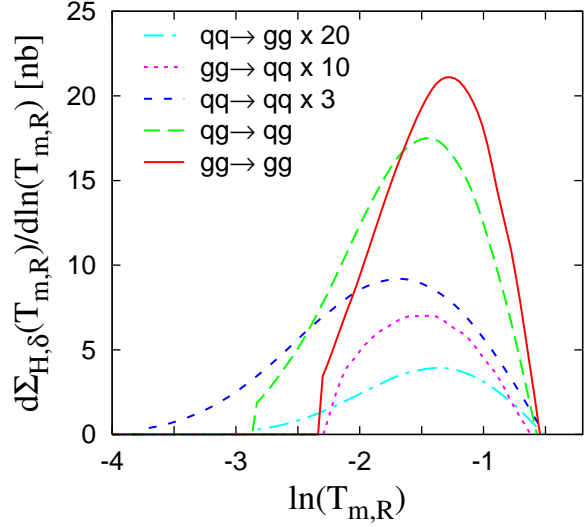}
\caption{The recoil thrust minor 
as predicted by CAESAR, with a cutoff before the divergence. Only a small fraction of the cross-section is beyond the cutoff.} 
\end{center}
\end{figure}

Below we present a table of the different event shapes mentioned here and the impact of the two main limitations we discussed, the beam-cut $\eta_{\rm{max}}$ and the breakdown of resummation due to divergences of $g_2$. Additionally we mention the expected impact of hadronisation corrections (not yet computed in full) on the different observables as well as the form of the 
estimated contribution from the underlying event. 

  \newcommand\TT{\rule{0pt}{3.3ex}}
  \newcommand\BB{\rule[-2.2ex]{0pt}{0pt}}

  {\small
  \begin{tabular}[t]{|c|c|c|c|c|} \hline
    Event-shape & Impact of $\eta_{\max}$ & \TT\BB
    \begin{minipage}[c]{0.18\tw}
      \begin{center}
        Resummation breakdown
      \end{center}
    \end{minipage}
    & 
    \begin{minipage}[c]{0.15\tw}
      \begin{center}
        Underlying Event
      \end{center}
    \end{minipage}
    & 
    \begin{minipage}[c]{0.17\tw}
      \begin{center}
        Jet hadronisation
      \end{center}
    \end{minipage}\\ \hline\hline
    $\tau_{\perp,g}$ & tolerable$^{*}$ & none & $\sim \eta_{\max}/Q$ & $\sim 1/Q$
    \\ \hline
    $T_{m,g}$ & tolerable & none& $\sim \eta_{\max}/Q$ & $\sim 1/(\sqrt{\as} Q)$
    \\ \hline
    $y_{23}$ & tolerable & none& $\sim \sqrt{y_{23}}/Q^*$ & $\sim \sqrt{y_{23}}/Q\, ^*$
    \\ \hline\hline
    $\tau_{\perp,\cE}$, $\rho_{X,\cE}$ & negligible & none& $\sim 1/Q$ & $\sim 1/Q$
    \\ \hline
    $B_{X,\cE}$ & negligible & none& $\sim 1/Q$ & $\sim
    1/(\sqrt{\as} Q)$
     \\ \hline\hline
    $T_{m,\cE}$ & negligible & serious& $\sim 1/Q$ & $\sim 1/(\sqrt{\as}Q)$
    \\ \hline
    $y_{23,\cE}$ & negligible & none& $\sim 1/Q$ & $\sim \sqrt{y_{23}}/Q \,^{*}$
    \\ \hline\hline
    $\tau_{\perp,\cR}$, $\rho_{X,\cR}$ & none & serious & $\sim 1/Q$ & $\sim 1/Q$
    \\ \hline
    $T_{m,\cR}$, $B_{X,\cR}$ & none & tolerable  & $\sim 1/Q$ & $\sim
    1/(\sqrt{\as} Q)$
    \\ \hline
    $y_{23,\cR}$ & none & intermediate$^{*}$ &  $\sim \sqrt{y_{23}}/Q \,^*$ & $\sim \sqrt{y_{23}}/Q \, ^*$
    \\ \hline
  \end{tabular}
  }

In the above the entries marked * are subject to uncertainty at present.

Further work is needed before the resummed expressions presented here can be compared with data including the matching to fixed order and computation of the power corrections for the various observables. This is currently in progress.

Having discussed the hadron-hadron event shapes we now move on to describe 
resummed studies concerning dijet production at HERA which can also be straightforwardly extended 
to hadron-hadron collisons.

\section{Dijet $p_t$ and angular spectra}
It has been known for some time that dijet total rates cannot be predicted within fixed-order QCD if symmetric cuts are applied to the two highest $p_t$ dijets \cite{FR,*KK}. While it was understood that the problems are 
to do with constraints on soft gluon emission, the exact nature of 
this constraint was only made 
clear in Ref.~\cite{BanDas}. There it was pointed out that there are large double logarithms (aside from single logarithms and less singular pieces) in the 
slope $\sigma'(\Delta)$  of the total rate, as a function of $\Delta$ the difference in minimum $p_t$ values of the two highest $p_t$ jets. These logarithms were resummed and it was shown that the slope of the total rate $\sigma' \to 0$ 
as  $\Delta \to 0$. This leads to a physical behaviour of the total rate as reflected by the data \cite{Zeus1}. 

To perform the comparison to data accurately however, requires two improvements to be made to the calculations of Ref.~\cite{BanDas}. Firstly the exact same jet algorithm has to be employed in the 
theoretical calculations and experimental measurements. The current algorithm 
used by H1 and ZEUS experiments is the inclusive $k_t$ algorithm. At hadron colliders variants of the cone algorithm are used and it is in fact a cone algorithm that was employed in Ref.~\cite{BanDas}. However the details of the calculation need to be ammended to define the cones in $\eta, \phi$ space as is done experimentally and calculations concerning this were presented at the working group meeting. The second important step is matching to fixed order estimates. 
We report below on the leading order matching to DISENT \cite{CatSey} while 
a full NLO matching is still awaited.

We also introduce and study two variables of related interest, the first is the 
difference in $p_t$, between the highest $p_t$ jets $\Delta p_{t,jj} 
= p_{t1}-p_{t2}$ (note that here we talk about the $p_t$ difference rather than 
the difference in the minimum $E_{\mathrm{cut}}$, that we mentioned earlier. 
The resummation of this distribution $\frac{d\sigma}{d\Delta p_{t,jj}}$ is essentially 
identical to that carried out in Ref.~\cite{BanDas}, except that here we 
compute the next-to--leading logarithms in different versions of the jet algorithm, which should help with direct experimental comparsions. We also perform the leading-order matching to DISENT.

Having developed the calculational techniques for $d\sigma/d{\Delta p_{t,jj}}$ 
it is then straightforward to generate the results for 
the distribution in azimuthal angle between jets $d\sigma/d\phi_{jj}$ which requires resummation in the region $\phi_{jj}=\pi$. These distributions have been measured at HERA and the Tevatron (most recently by the D0 collaboration).
Comparing the resummation with data would represent an interesting challenge for the theory insofar as the status of resummation tools is concerned, 
and is potentially very instructive. 

\subsection{The $\Delta p_{t,jj}$ and $\phi_{jj}$  distributions}
We shall consider dijet production in the DIS Breit frame. 
For the jet definition we can consider either an $\eta,\phi$ cone algorithm 
(such as the infrared and collinear safe midpoint cone algorithm) or the inclusive $k_t$ algorithm. We shall point out to what level the two algorithms would give the same result and where they can be expected to differ. We shall use a four-vector recombination scheme where the jet four-momentum is the sum of individual constituent hadron four-momenta.
We also impose cuts on the highest $p_t$ jets such that $|\eta_{1,2}| \leq 1$ and $p_{t1,t2}\geq E_{\mathrm{min}}$.

We then consider the quantity $\Delta p_{t,jj}=p_{t1}-p_{t2}$ which vanishes at Born order and hence the distribution at this order is just $\frac{d\sigma}{dp_{t,jj}} \propto \delta(p_{t,jj})$.

Beyond leading order the kinematical situation in the plane normal to the Breit axis is represented as before \cite{BanDas}: 
\begin{eqnarray}
\vec{p_{t1}} &=& p_{t1} (1,0) \\
\vec{p_{t2}} &=& p_{t2} \left (\cos (\pi \pm \epsilon), \sin(\pi \pm \epsilon) \right )\\
\vec{k_t}    &=& k_t \left(\cos \phi, \sin \phi\right)
\end{eqnarray}

Thus we are considering a small deviation from the Born configuration of jets back-to--back in azimuth, induced by the presence of a soft gluon with transverse momentum $k_t \ll p_{t1,t2} $ (which is not recombined by the algorithm with either hard parton) and with azimuthal angle $\phi$. In the above $\epsilon$ represents the recoil angle due to soft emission. 
We then have 
\begin{equation}
\Delta p_{t,jj}= |p_{t1}-p_{t2}| \approx |k_t \cos \phi|, 
\end{equation}
which accounts for the recoil $\epsilon$ to first order and 
hence is correct to NLL accuracy. Thus for the emission of several soft gluons 
we have the $p_t$ mismatch given by 
\begin{equation}
\label{pspace}
\Delta p_{t,jj}=|\sum_{i \notin j} k_{xi}|, 
\end{equation}
where $k_x$ denotes the 
single component of gluon transverse momentum, along the direction of the hard jets, which are nearly back-to--back in the transverse plane. The sum includes only partons not merged by the algorithm into the highest $E_t$ jets.

Similarly for the dijet azimuthal angle distribution\footnote{Note that the kinematical relations we derive here would be equally valid for dijets produced in hadron-hadron collisions at the Tevatron or LHC and just the dynamics of multisoft gluon emission would be more complex.}, we have : 
\begin{equation}
\pi-\phi_{jj} \approx \frac{1}{p_t} |\sum_{i \notin j} k_{yi} |.
\end{equation}
where 
$\phi_{jj}$ is the azimuthal angle between the two 
highest $p_t$ jets. Note that in the above we have set $p_{t1}=p_{t2}=p_t$ since we are considering a small deviation from the Born configuration and this approximation is correct to NLL accuracy. We also introduced $k_y$, the component of 
soft gluon momentum normal to the jet axis in the transverse plane.

In either of the above two cases, i.e the $\Delta p_{t,jj}$ or $\phi_{jj}$ distributions, an identical resummation is involved, due to the similar role of soft partons not recombined into jets.  
Henceforth we shall proceed with just the $\Delta p_{t,jj}$ resummation results, it being understood that similar considerations apply to $\phi_{jj}$ in the region $\phi_{jj} \sim \pi$.

Assuming independent emission of soft gluons by the hard three-parton system 
(the incoming parton and the two outgoing partons that initiate the dijets) and factorising the phase-space Eq.~\ref{pspace} as below\footnote{We compute here the cross-section for the observable to be less than $\Delta p_{t,jj}$ from which we can easily obtain the corresponding distribution.}:
\begin{equation}
\label{constr}
\Theta \left (\Delta p_{t,jj}-|\sum _{i \notin j} k_{x,i}|\right) = \frac{1}{\pi} \int_{-\infty}^{\infty} \frac{db}{b} \sin(b \Delta p_{t,jj}) \prod_{i \notin j} e^{ib k_{xi}},
\end{equation}
the resummed result for the $\Delta p_{t,jj}$ distribution can be expressed as 
\begin{equation}
\label{eq:sigmadif}
\frac{d^3\sigma}{dx dQ^2 d\Delta p_{t,jj}}(E_\mathrm{{min}},\Delta p_{t,jj}) = 
\sum_{\delta=q,g} 
\int_x^1 \frac{d\xi}{\xi} \int_0^1 dz \sum_{a=T,L} F_a (y) C^a_\delta(\xi,z,E_{\mathrm{min}}) w_\delta(Q,\Delta p_{t,jj}).
\end{equation}
In the above $\xi$ and $z$ are phase-space variables that parametrise the 
Born dijet configuration, $F_{a=T,L}$  denotes the $y =Q^2/xs$ dependence associated to the transverse or longitudinal structure function while $C^a$ is the Born matrix-element squared.
The function $w$ represents the result of resummation.

The resummed expression $w$ requires some explanation. Its form is as follows
\begin{equation}
w_\delta(p_{t,jj}) = \int_0^{\infty} \frac{db}{b} \sin(b \Delta p_{t,jj}) \exp[-R_{\delta} (b)] {\mathcal{S}}(b) q_\delta \left (x/\xi,1/b^2 \right).
\end{equation} 
Note the fact that the exponentiation holds only in $b$ space where $b$ is the impact parameter. The function $R(b)$ (we ignore the subscript $\delta$ which describes either incoming quarks or gluons) is the Sudakov exponent which can be computed up to NLL accuracy,
\begin{equation}
R(b) = L g_1(\alpha_sL)+g_2(\alpha_sL), \,\, L \sim\ln(bQ).
\end{equation}
while $S(b)$ is the non-global contribution that arises from soft 
partons inside the jet emitting outside it. $q_\delta$ is the incoming quark or gluon density and its scale depends on the variable $b$. The functions $g_1$ and $g_2$ are the leading-logarithmic and next-to--leading logarithmic resummed quantities.

For the leading logarithms $g_1$ and a subset of 
next-to--leading logarithms $g_2$, 
generated essentially by exponentiation of the single-log result in $b$ space, the cone and inclusive $k_t$ algorithms would give the same result, 
which we have computed. Starting from terms that begin with 
$\alpha_s^2 \ln^ 2 b$ in $g_2$ (specifically two soft wide-angle gluons), 
the following two effects become important:
\begin{itemize}
\item For cone algorithms the implementation of the split/merge stage affects the $g_2$ piece. Present calculations \cite{BanDas} are valid to NLL accuracy 
if {\emph{all}} 
the energy shared by overlapping jets is given to the jet that would have highest $p_t$.  Note that this is different from merging the overlapping jets themselves. If other merging procedures are used the calculation becomes more complex but is still tractable.
\item For the $k_t$ algorithm it is just being realised that running the 
algorithm generates terms that start at $\alpha_s^2 \ln^2b$ in the exponent, which are not correctly treated by naive Sudakov exponentiation. 
These terms, which are generated by the clustering procedure, 
can also be numerically accounted for in our case, but this is work in progress.
\end{itemize}
The effects that we mention above 
cause a similar impact on the final result as the non-global term $S(b)$ which was shown to be at around the 10\% level in Ref.~\cite{BanDas}. Hence the current results for the $k_t$ algorithm 
that do not account for the recently found additional terms and only approximately for the non-global logs, can be expected to change by around 10\% when these effects will be included correctly. 

\begin{figure}
\begin{center}
\epsfig{file=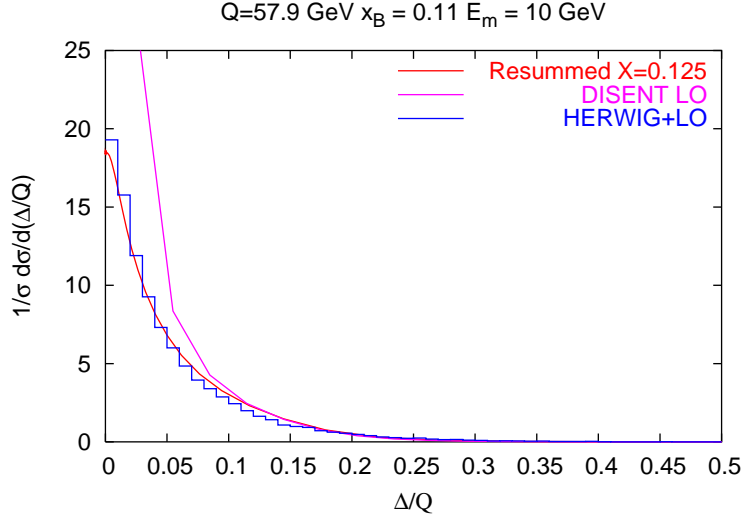}
\label{etdifmat}
\caption{Figure showing the resummed result matched to fixed-order DISENT results for the variable $\Delta = Q \Delta{p_{t,jj}}$. Also shown, for comparison, are HERWIG results with matrix-element corrections and the DISENT result alone.}
\end{center}
\end{figure}

We present in Fig.~6 preliminary 
results for the $\Delta p_{t,jj}$ distribution 
matched to the leading order DISENT prediction, using the $k_t$ algorithm. 
The matching at present combines quark and gluon channels wheras 
ideally one would like to separate the incoming quark and gluon channels 
with the right weights (${\mathcal{O}}(\alpha_s)$ coefficient functions). 
This would be possible if, for instance, there was parton 
flavour information explicit in the fixed order codes, a limitation of the fixed-order codes that needs to be addressed also for hadron-hadron event shapes to be matched to NLO predictions. 

We also present a comparison with HERWIG \cite{HERWIG} results on the same quantity. The variable $X$ in the figure merely refers to the effect of 
using the jet $p_t$ as the hard scale rather than the photon virtuality $Q^2$, formally a NNLL effect. 
It is amusing to note the very good agreement of the resummation 
with HERWIG but not too much can be read into it at this stage. Given the minor role of non-global effects we would expect HERWIG and our predictions to indeed have a broad resemblence. However we should mention that the resummed result in Fig.~6 is at present subject to change pending proper inclusion of non-global logs and the effect of independent soft emission at large angles. The latter is partly included in the results shown, 
through exponentiation of the one-gluon result as we pointed out before, but the clustering procedure changes this result at about the same level as the non-global logs (${\mathcal{O}}(\alpha_s^2 \ln^2b)$ in the exponent), and this feature needs to be accounted for still. 
Secondly the matching to LO DISENT combines channels and this spoils control over the 
$\alpha_s^2 \ln^2 Q/\Delta p_{t,jj}$ term in the expansion of the resummation to NLO. A full NLO matching with proper separation of the channels is awaited. The HERWIG curve 
also includes an intrinsic $k_t$ component that lowers the height of the result at small $p_{t,jj}$, which can be easily 
included in the theoretical resummation but at present is excluded. 
Given these differences the very good agreement one sees with HERWIG is expected to 
change to some extent although broadly speaking the shapes of the two curves are expected to be similar.
Similar conclusions apply for the $\phi_{jj}$ observable. 
\section{The vector $Q_t$ of the current hemisphere}
Next we examine a quantity that, as mentioned in the introduction, makes a very good analogy with Drell-Yan transverse momentum, $Q_t$, distributions. Comparison of the resummation of this observable with data could help to understand whether extra broadening of conventionally resummed 
$Q_t$ spectra, is generated at small $x$. If so this will be a significant factor at the LHC. The observable in question is the (modulus of) the vectorially summed transverse momenta of all particles in the Breit frame current hemisphere:
\begin{equation}
Q_t = |\sum_{i \in {\mathcal{H}}_c} \vec{k}_{t,i}|.
\end{equation}

Using momentum conservation this quantity is simply equal to the modulus of the transverse momenta of emissions in the remnant hemisphere. These emissions can all be ascribed to the incoming quark to NLL accuracy, {\emph{apart}} from the soft wide-angle component where large-angle emissions in the current hemisphere can emit softer gluons into the remnant hemisphere (the by now familiar non-global logarithms).

The resummed result for this observable can be expressed as :
\begin{equation}
\frac{d\sigma}{dQ_T^2} \sim
\sigma_0 \int_{0}^{\infty} bdb J_0(bQ_t) \exp[-R(b)] S(b) q(x,1/b^2)
\end{equation}
where $J_0$ is the zeroth order Bessel function, $R(b)$ is the Sudakov exponent (the ``radiator'') , $S(b)$ the non-global contribution and $q$ denotes the quark distribution summed over quark flavours with appropriate weights (charges).

The result for the radiator to NLL accuracy can be expressed, as before, in terms of a leading-log and next-to--leading log function:
\begin{equation}
R(b) = Lg_1(\alpha_s L) + g_2 (\alpha_s L), \,  L = \ln(bQ).
\end{equation}
We have 
\begin{eqnarray}
g_1 &=& \frac{C_F}{2 \pi \beta_0 \lambda} [-\lambda-\ln(1-\lambda)], \\ 
g_2 &=& \frac{3 C_F}{4 \pi \beta_0} \ln (1-\lambda)+\frac{K C_F}{4 \pi^2 \beta_0^2} \left[\frac{\lambda}{1-\lambda}+\ln(1-\lambda)\right] \\ \nonumber
    &+&\frac{C_F}{2\pi} \left (\frac{\beta_1}{\beta_0^3}\right)\left 
[-\frac{1}{2} \ln^2 \left (1-\lambda \right)-\frac{\lambda+(1-\lambda)}{1-\lambda} \right ],
\end{eqnarray}
where we have
$\lambda = \beta_0 \alpha_s \ln [Q^2 (\bar{b})^2], \, \,\bar{b} = be^{\gamma_E}/2$ and $K = (67/18-\pi^2/6)C_A -5/9 \, n_f$.

It is straightforward to express the result directly in $Q_t$ space and one has for the pure NLL resummed terms:
\begin{equation}
\label{qtresum}
\frac{d\sigma}{dQ_T^2} \sim \frac{d}{dQ_T^2} \left[e^{-R(Q/Q_t)-\gamma_E R'(Q/Q_t)} \frac{\Gamma\left(1-R'/2\right)}{\Gamma\left(1+R'/2\right)} q(x,Q_T^2)S(Q/Q_t)\right]
\end{equation}
where $R'=dR/d\ln(Q/Q_t)$. The result has a divergence at $R'=2$ which is due to retaining just NLL terms and is of the same nature as that 
discussed before 
for certain hadron-hadron event shapes and the Drell-Yan $Q_t$ distribution. 
However in the present case the divergence is at quite low values of $Q_t$, e.g for $Q=100$ GeV, the divergence is at around $0.5$ GeV (depending on the exact choice for $\Lambda_{\rm{QCD}}$). Thus it is possible to safely study the distribution down to $Q_t$ values of a few GeV using the simple form Eq.~\ref{qtresum}.
We note that is is also possible to eliminate the divergence if one defines the radiator such that $R(b)\to R(b) \theta(\bar{b}Q-1)$, which is a restriction that follows from leading-order kinematics (that one assumes to hold at all orders).
The resultant modification has only a negligible impact in the $Q_t$ range that we expect to study phenomenologically.

After the matching to fixed-order is performed, we can probe the 
non-perturbative smearing $e^{-g b^2}$ that one can apply to the $b$ space resummed result. Comparisons with data should hopefully reveal whether the NLL resummed result + `intrinsic $k_t$' smearing, mentioned above, is sufficient at smaller values of $x$ or whether extra broadening is generated in the small $x$ region, that has a significant effect on the result. Data from H1 are 
already available for this distribution \cite{Kluge} and this should enable rapid developments concerning the above issue.

\section{Conclusions}
In this article we have provided a summary of the developments discussed at the HERA-LHC workshop working group 2, concerning the topic of all-order QCD resummations. Specifically we have mentioned recent work 
carried out for hadronic dijet event shapes, dijet $E_t$ and angular spectra and resummation of the current-hemisphere transverse momentum distribution in the DIS Breit frame. 

We have stressed the important role of HERA studies in the development of the subject from the LEP era and the fact that, in this regard, HERA has acted as a bridge between LEP studies of the past (although LEP analysis of data continues and is an important source of information) and future studies at both 
the Tevatron and the LHC. 

We have particularly tried to stress the continuing crucial role of HERA in testing all-order QCD dynamics, especially in the context of multi-hard parton observables where studies are currently ongoing. Careful experimental and theoretical collaborative effort is 
needed here in order to confirm the picture developed for 
NLL resummations and power corrections. If this program is successful 
it will greatly ease the way for accurate QCD 
studies at more complex hadronic environments, such as the LHC.

\section*{Acknowledgements}
We would like to thank the convenors and organisers of the series of meetings 
which were a part of the HERA-LHC workshop, for their skillful organisation 
and for providing us all the necessary facilities needed to present and 
develop our respective contributions here.



\bibliographystyle{heralhc} 
{\raggedright
\bibliography{heralhc}
}
\end{document}